\documentclass[twocolumn,pra,unsortedaddress,showpacs,floatfix,
citeautoscript,nofootinbib]{revtex4-1}
\usepackage{amsmath,amssymb,mathrsfs}
\DeclareMathOperator{\sech}{sech}
\usepackage{amsbsy}
\usepackage{psfrag}
\usepackage{graphicx}
\usepackage{epsfig}
\usepackage{bm}
\usepackage{color}

\usepackage[normalem]{ulem}
\usepackage{latexsym}
\usepackage[colorlinks,urlcolor=blue,citecolor=blue]{hyperref}
\usepackage{dcolumn}
\usepackage[normalem]{ulem}
\usepackage{subfigure}

\begin{document}

\title [Equation of state of the one- and three-dimensional Bose-Bose gases] 
{Equation of state of the one- and three-dimensional Bose-Bose gases}
\author{Emerson Chiquillo}
\address{Escuela de F\'isica, Universidad Pedag\'ogica y Tecnol\'ogica de 
Colombia (UPTC),\\
Avenida Central del Norte, 150003 Tunja, Colombia}

\begin{abstract}
We calculate the equation of state of Bose-Bose gases in one and 
three dimensions in the framework of an effective quantum field theory.
The beyond-mean-field approximation at zero-temperature and the one-loop 
finite-temperature results are obtained performing functional integration
on a local effective action.
The ultraviolet divergent zero-point quantum fluctuations are removed by 
means of dimensional regularization.
We derive the nonlinear Schr\"odinger equation to describe one- and 
three-dimensional Bose-Bose mixtures and solve it analytically in the
one-dimensional scenario.
This equation supports self-trapped brightlike solitonic-droplets and 
self-trapped darklike solitons.
At low temperature, we also find that the pressure and the number of 
particles of symmetric quantum droplets have a nontrivial dependence 
on the chemical potential and the difference between the intra- and the
inter-species coupling constants.
\end{abstract}

\maketitle

\section{Introduction}
The Bose gases at zero temperature have been studied extensively for over
60 years. The main tool in this study lies in the use of the mean-field 
Gross-Pitaevskii equation (GPE).
The leading quantum corrections to the mean-field results for the
three-dimensional (3D) Bose gas were calculated by Lee, Yang, and Huang 
by using the pseudopotential method \cite{LHY,beyond1}.
In the last years, some experiments with 3D ultracold and dilute atomic
gases \cite{LHY-3D} have put in evidence beyond-mean-field
effects in the equation of state of repulsive bosons.
The quantum depletion of an interacting homogeneous 3D 
Bose-Einstein condensate (BEC) is measured in good agreement with the 
Bogoliubov theoretical results \cite{Bogol-exper}.
A nonperturbative renormalization-group approach is employed in the study
of a quasi-two-dimensional (quasi-2D) Bose gas with or without an optical 
lattice \cite{optical-lattice}. Both the chemical potential and the
interaction constant dependencies are in good agreement with experimental
data.
Recently, the formalism of functional integration at zero and finite
temperature is consider to derive the equation of state in one-, two-,
and three-dimensional ultracold single-component Bose gases
\cite{Regularization}. This work shows that the ultraviolet divergent
zero-point energy arising in the Gaussian fluctuations is removed by 
dimensional regularization.
The functional integration is also used to investigate the nonuniversal
corrections to the equation of state in three-dimensional Bose gases
\cite{Finite-range}. The highly nontrivial two-dimensional (2D) Bose gas
is investigated in Ref. \cite{2D}.
Even at zero temperature, the first quantum correction beyond-mean-field
approximation of Bose gases with magnetic atoms has revealed the crucial
role played by the quantum fluctuations both experimentally 
\cite{Dipolar1} and theoretically \cite{Dipolar2}. 

At present, Bose-Bose mixtures at zero temperature in one, two, and 
three dimensions have attracted great interest \cite{Petrov1,Petrov2}.
In the study of these mixtures the competition between the mean-field term
and the beyond-mean-field correction leads to the possibility of the 
formation of a droplet phase.
The superfluidity of two weakly interacting Bose gases at finite
temperature also has been studied in the frame of the Bogoliubov
model \cite{mixture1}.
In a mixture of two interacting repulsive Bose condensates in a 
quasi-one-dimensional (quasi-1D) geometry, a finite-temperature 
path-integral Monte Carlo method is employed to study the effect of both 
temperature and inter-species interactions on the density profile and the
superfluid fraction \cite{Montecarlo}.
The hydrodynamic modes of partially condensed Bose mixtures within Popov's
approximation are studied in \cite{Modes}.
The investigation of the effects of quantum fluctuations in a 3D Rabi 
coupled two-component Bose gas of interacting alkali atoms is performed in
\cite{Rabi}.
Also it is important to stress the experimental achievements with mixtures
of ultracold and dilute gases of $^{39}$K atoms in two hyperfine states,
$|F=1; m_{F} = -1,0 \rangle $.
These experimental breakthroughs have put in evidence beyond-mean-field
effects on the equation of state of Bose-Bose mixtures.
The observation of quantum droplets in a 1D lattice is reported in
\cite{Exper-droplet1}.
Self-bound droplets in free space solely stabilized by the contact
interactions are obtained in \cite{Exper-droplet2}.
An attractive mixture of BECs confined in an optical waveguide shows two 
types of self-bound states, bright solitons and quantum liquid droplets
\cite{Exper-droplet3}.

In this work we investigate the effect of the quantum Gaussian
fluctuations at zero and finite temperature in 1D and 3D Bose-Bose 
mixtures through an effective quantum field theory.
We derive the equation of state for both $D$-dimensional
Bose-Bose mixtures by means of functional integration. We include dimensional 
regularization of the zero-point energy.
Interestingly, we find that within a local density description, the 
nonlinear Schr\"odinger equation (NLSE) to describe 1D free Bose-Bose
mixtures has analytical spatially localized solutions in the form of 
self-bound brightlike solitonic droplets and self-bound darklike 
solitons.
Our finite-temperature results generalize some those obtained at 
zero-temperature by using suitable scattering arguments in the regime of 
perturbative expansions \cite{Petrov1,Petrov2}.
We also find an extension of the equation of state of single-component 
Bose gases \cite{Regularization}.

\section{$D$-dimensional functional integration of the Bose-Bose mixtures}
We investigate the interacting dilute and ultracold Bose-Bose gases with 
two relevant hyperfine states $(\uparrow,\downarrow)$, in $D$ spatial 
dimensions ($D$= 1,3), at temperature $T$, equal-mass $m$, and chemical
potentials $\mu_{\uparrow},\mu_{\downarrow}$.
In this study we use the path-integral formalism, where 
each component is described by a complex bosonic field
$\psi_\alpha$ ($\alpha=\uparrow$,$\downarrow$). 
So, given the spinor $\Psi=(\psi_\uparrow,\psi_\downarrow)^T$, the
Euclidean Lagrangian density in a $D$-dimensional box of volume $L^D$ is 
given by 
\begin{eqnarray} 
\mathscr{L}[\Psi,\Psi^*] &=& \sum_{\alpha=\uparrow,\downarrow} 
\Big[\psi_\alpha ^*\Big(\hbar\frac{\partial}{\partial\tau} 
- \frac{\hbar^2}{2m} \nabla^2  - \mu_\alpha\Big)\psi_\alpha 
\nonumber \\
&+& \frac{1}{2}\sum_{\sigma=\uparrow,\downarrow}g_{\alpha\sigma}
|\psi_\alpha|^2|\psi_\sigma|^2\Big],
\end{eqnarray}
where the fields are considered at the position $\mathbf{r}$ and the
imaginary time $\tau$. Here, we consider a repulsive mixture where 
$g_{\alpha\alpha},g_{\alpha\sigma}>0$ are the strengths of the 
intra- and the inter-species coupling constants, respectively.
These couplings are related to the $s$-wave scattering lengths 
$a_{\alpha\alpha},a_{\alpha\sigma}$ for collisions between the components
$\alpha$ and $\sigma$ in the universal regime.
In this regime, the interaction effects between atoms could be estimated 
by means of the $s$-wave scattering length. Thus all the interatomic
potentials with the same $s$-wave scattering length will have the same
properties to leading order in the low-density expansion 
\cite{Andersen,Universal,Pit-Strin}. 
At higher order in the low-density expansion, the properties will depend
on the details of the interatomic potential and we have nonuniversal 
effects. These $s$-wave scattering lengths depend on the dimensionality of
the system as it is discussed later.
Thermodynamic properties of the system can be obtained from the grand 
canonical partition function
$\mathcal{Z}
=\int \mathcal{D}[\Psi,\Psi^*]\exp\big(-{S[\Psi,\Psi^*]/\hbar}
\big)$, where the action is given by
$S[\Psi,\Psi^*]=\int_0^{\hbar \beta}d\tau \int_{L^D} d^D r 
\mathscr{L}[\Psi,\Psi^*]$ and $\beta\equiv 1/k_B T$, with $k_B$ the
Boltzmann's constant.
We consider the superfluid phase, where a U(1) gauge symmetry of each 
component is spontaneously broken \cite{Rabi}.
Then we can set $\psi_\alpha (\textbf{r},\tau) = \sqrt{n_\alpha} 
+ \eta_\alpha (\textbf{r},\tau)$, where $\sqrt{n_\alpha}$ corresponds to 
the mean-field approximation, with $n_{\alpha}=|\psi_{\alpha}|^2$ the
three-dimensional density of the condensate or one- and two-dimensional 
quasicondensate density \cite{quasi,Pit-Strin}. 
The Gaussian fluctuations (one loop) around $\sqrt{n_\alpha}$ are given
by $\eta_\alpha (\textbf{r},\tau)$.
In order to calculate the ground state of the $D$-dimensional mixtures, we
expand the action up to the second order in 
$\eta_\alpha (\textbf{r},\tau)$ and $\eta^*_\alpha (\textbf{r},\tau)$
\cite{Modes}.
Then, the grand potential $\Omega = -\beta^{-1} \ln \mathcal{Z}$, can be
split as $  \Omega(\mu_\uparrow,\mu_\downarrow,\sqrt{n_\uparrow},
\sqrt{n_\downarrow},T)=
\Omega_0(\mu_\uparrow,\mu_\downarrow,\sqrt{n_\uparrow}, 
\sqrt{n_\downarrow})
+ \Omega_g(\mu_\uparrow,\mu_\downarrow,
\sqrt{n_\uparrow},\sqrt{n_\downarrow},T)$,
where $\Omega_0$ is the mean-field contribution, while $\Omega_g$ takes 
into account Gaussian fluctuations.

\section{Mean-field approximation}
Using the Bogoliubov approximation, we consider 1D and 3D Bose-Bose
mixtures, where
\begin{eqnarray} 
\frac{\Omega_0}{L^D}
= \sum_{\alpha=\uparrow,\downarrow} \Big(-\mu_\alpha n_\alpha
+ \frac{1}{2}\sum_{\sigma=\uparrow,\downarrow}g_{\alpha\sigma}
n_\alpha n_\sigma\Big).
\label{mean-field}
\end{eqnarray}
Since we want $\sqrt{n_\alpha}$ to describe the 3D BEC or quasi-1D BEC, 
then in the action, the linear terms in the fluctuations vanish such that
$\sqrt{n_\alpha}$ really minimizes the action \cite{Stoof}. 
Then the mean-field approximation is obtained by minimizing $\Omega_0$, 
namely $\partial\Omega_0/\partial\sqrt{n_\alpha}=0$ \cite{Regularization},
so
\begin{eqnarray} 
\mu_\alpha = g_{\alpha\alpha}n_\alpha + g_{\alpha\lambda}n_\lambda 
\label{chem}
\end{eqnarray}
with $\alpha,\lambda=\uparrow,\downarrow$ and $\alpha\neq \lambda$.
Thus, taking into account the relation $P_0= - \Omega_0/L^D$, the
mean-field equation of state is given by
\begin{eqnarray} 
P_0 (\mu_{\uparrow},\mu_{\downarrow}) = \frac{1}{2}\sum_{\alpha}
\frac{g_{\alpha\alpha}\mu_{\lambda}^2
- g_{\alpha\lambda}\mu_{\alpha}\mu_{\lambda}}
{g_{\alpha\alpha}g_{\lambda\lambda} -g_{\alpha\lambda}^2}
\label{Pressure0}
\end{eqnarray}
with $\alpha,\lambda=\uparrow,\downarrow$ and $\alpha\neq \lambda$.
In the low-energy limit of the two-body scattering, the Lippman-Schwinger 
equation renormalizes the coupling constant of the contact interaction
between particles \cite{Regularization,Stoof}. In this theory we have that 
for the 3D Bose-Bose gases the strength of the interaction can be written
as
\begin{eqnarray} 
g_{\alpha\sigma}\equiv\frac{4\pi\hbar^2a_{\alpha\sigma}}{m}.
\label{g3D}
\end{eqnarray}
The solution of the one-dimensional atom-atom scattering problem within
the pseudopotential approximation provides an effective 1D coupling 
constant for atoms in the presence of transverse harmonic confinement with
trapping frequency $\omega_{\perp}$, and length of the transverse trap
$l_{\perp} = \sqrt{\hbar/(m\omega_{\perp})}$ \cite{scattering1D}.
Here we consider an extension to 1D Bose-Bose gases where 
$g_{\alpha\sigma}$ represents the effective 1D coupling constant. This
can be read as \cite{scattering1D,Pit-Strin}
\begin{eqnarray} 
g_{\alpha\sigma}\equiv \frac{2\hbar^2}{ml_\perp ^2}
\frac{a_{\alpha\sigma}}{1-Ca_{\alpha\sigma}/l_\perp},
\label{g1D}
\end{eqnarray}
where $C=-\zeta(1/2)/\sqrt{2} \simeq 1.0326$, with $\zeta(x)$ denoting the
Riemann's zeta function.
It is worth noting that the transversal confinement induces a resonant 
behavior of the 1D coupling constant called confinement-induced resonance
(CIR). This occurs when $a_{\alpha\sigma} \sim 0.968l_\perp$. So, an
effective and repulsive 1D potential is only achieved in the range
$0<a_{\alpha\sigma}<0.968l_\perp$.
On the other hand, a simple 1D coupling constant is obtained in the 1D 
mean-field regime \cite{scattering1D,Pit-Strin}. 
As long as the radial confinement, fixed by the radial oscillator length, 
is much greater than the scattering length, 
$l_{\perp} \gg a_{\alpha\sigma}$, the transverse ground state can be 
modeled with a Gaussian profile and the coupling constant takes the form 
$g_{\alpha\sigma} = 2\hbar^2 a_{\alpha\sigma} / ml_\perp ^2$.
The effective and repulsive 1D potential is reached in the
semi-infinite range $a_{\alpha\sigma}>0$.
This mean-field result holds under the condition
$l_{\perp}/a_{\alpha\sigma}\gg 1$ on the renormalized and effective 1D 
coupling constant given by Eq. (\ref{g1D}).

\section{Beyond-mean-field corrections}
The grand potential of the Gaussian fluctuations 
$\Omega_g (\mu_{\uparrow},\mu_{\downarrow},T)$ is provided by
\begin{eqnarray} 
\Omega_g = - \frac{1}{2\beta}\sum^{+\infty}_{\substack{\textbf{k}>0\\
n=-\infty}} \ln \bigg[(\hbar^2\omega_n ^2 + E_+^2)
(\hbar^2\omega_n ^2 + E^2_-)\bigg],
\label{Grand1}
\end{eqnarray}
with the bosonic Matsubara's frequencies $\omega_n=2\pi n/\hbar\beta$, and
the Bogoliubov's modes 
$E_\pm (k,\mu_{\uparrow},\mu_{\downarrow}) = [{\varepsilon^2(k)} 
+ 2{\varepsilon(k)f^2_{\pm}(\mu_{\uparrow},\mu_{\downarrow})}]^{1/2}$.
Here, we have the free-particle energy $\varepsilon(k)= \hbar^2 k^2/2m$,
and the function $f_{\pm}(\mu_{\uparrow},\mu_{\downarrow})$ is defined by
\begin{eqnarray} 
f^2_{\pm} = \frac{\xi}{2} \Big[A + B &\pm& \sqrt{(A-B)^2 
+ 4\epsilon AB}\Big],
\label{*sound}
\end{eqnarray}
where $\xi=[g_{\uparrow\uparrow}g_{\downarrow\downarrow}
(1-\epsilon)]^{-1}$, 
$\epsilon=
g_{\uparrow\downarrow}^2/g_{\uparrow\uparrow}g_{\downarrow\downarrow}$,
$A=g_{\uparrow\uparrow} (g_{\downarrow\downarrow}\mu_{\uparrow} 
-g_{\uparrow\downarrow}\mu_{\downarrow})$, and $B=g_{\downarrow\downarrow}
(g_{\uparrow\uparrow}\mu_{\downarrow} 
- g_{\uparrow\downarrow}\mu_{\uparrow})$.
The sum over the bosonic Matsubara's frequencies given by 
Eq. (\ref{Grand1}) can be read as \cite{Le Bellac}
\begin{eqnarray} 
\Omega_g =\frac{1}{2}\sum_{k,\pm}\Big[E_{\pm}
+ \frac{2}{\beta} \ln \big(1-e^{-\beta E_\pm }\big)\Big].
\label{Grand2}
\end{eqnarray}
Thus, in the grand potential, the contribution to the quantum Gaussian 
fluctuations can be written as $ \Omega_g(\mu_\uparrow,\mu_\downarrow,T)=
\Omega_g^{(0)}(\mu_\uparrow,\mu_\downarrow)
+\Omega_g^{(T)}(\mu_\uparrow,\mu_\downarrow)$,
where $\Omega_g^{(0)}$ is given by the first term in Eq. (\ref{Grand2}),
and it represents the zero-temperature contribution to the fluctuations.
$\Omega_g^{(T)}$ is given by the second term in Eq. (\ref{Grand2}),
and it considers the thermal Gaussian fluctuations.
This last contribution to the grand potential is tantamount to that of
the microscopic Popov's theory \cite{Modes}, employed in the study of
the hydrodynamic modes of partially condensed two-component Bose 
mixtures. However, it is worth noting that in that work, the first term 
given by our Eq. (\ref{Grand2}) is not considered. This is key to
explaining the existence of zero-temperature quantum droplets 
\cite{Petrov1,Petrov2}.
In the continuum limit, where $\sum_k\rightarrow L^D \int d^Dk/(2\pi)^D$,
the zero-point grand potential $\Omega_g^{(0)}$ is ultraviolet
divergent at any integer dimension $D$, namely, $D=1,2,3$. 
However, this divergence can be eliminated by dimensional regularization 
\cite{Hooft,Andersen,Regularization}. 
Thus, the zero-point grand potential of a repulsive $D$-dimensional
Bose-Bose mixture can be written as
\begin{eqnarray}
\Omega_g^{(0)}(\mu_{\uparrow},\mu_{\downarrow})
= \gamma_D \sum_{\pm}f_{\pm}^{D+2}(\mu_{\uparrow},\mu_{\downarrow}),
\label{regularization}
\end{eqnarray}
where 
\begin{eqnarray} 
\gamma_D = \frac{L^D}{\Gamma(D/2)}\Big(\frac{m}{\pi \hbar^2}\Big)^{D/2}
B\Big(\frac{D+1}{2}, -\frac{D+2}{2}\Big) 
\end{eqnarray}
with the Euler's Beta function $B(x,y)=\Gamma(x)\Gamma(y)/\Gamma(x+y)$,
and $\Gamma(x)$ the Euler's Gamma function.
It is worth mentioning that the regularization of the 2D Bose-Bose gases
requires a more elaborated treatment.
For a single-component 2D Bose gas, the two-dimensional integral is
extended to a noninteger $D=2 - \varepsilon$ dimension, and the limit
$\varepsilon \rightarrow 0$ is applied at the end of the calculation
\cite{Regularization,2D}. 
For the 1D Bose-Bose mixtures we have an attractive term, 
$ \gamma_1 = - (2L/3\pi)(m/\hbar^2)^{1/2} $,
while for the 3D Bose-Bose mixtures we have a repulsive one,
$\gamma_3 = (8L^3/15\pi^2)(m/\hbar^2)^{3/2}$.
Thus, taking into account the relation $P^{(0)}= - \Omega_g ^{(0)}/L^D$, 
we can easily obtain the pressure of one- and three-dimensional Bose-Bose
gases. So, at zero temperature, the equation of state in the universal regime
beyond-mean-field approximation, or the Lee-Huang-Yang (LHY) contribution, 
can be read as
\begin{eqnarray} 
P^{(0)}(\mu_{\uparrow},\mu_{\downarrow}) 
= -\frac{\gamma_D}{L^D} 
\sum_{\pm}f_{\pm}^{D+2}(\mu_{\uparrow},\mu_{\downarrow}).
\label{Pressure1}
\end{eqnarray}
Now, using the mean-field term given by Eq. (\ref{mean-field}), the
coupling constants (\ref{g3D}) and (\ref{g1D}), and the grand potential 
provided by the Eq. (\ref{regularization}), we get an explicit form of the
energy of the $D$-dimensional Bose-Bose gases including the LHY term
\begin{eqnarray} 
\frac{E_D}{L^D} = P_0 + \frac{\gamma_D}{L^D} \sum_{\pm}f_{\pm}^{D+2}.
\end{eqnarray}
From relation (\ref{chem}), we also obtain the energy in terms of the 
density \cite{Petrov1,Petrov2},
\begin{eqnarray} 
\frac{E_D}{L^D}
= \frac{1}{2}\sum_{\alpha,\sigma=\uparrow,\downarrow}
g_{\alpha\sigma} n_\alpha n_\sigma
+ \frac{\gamma_D}{L^D} \sum_{\pm}c_{\pm}^{D+2},
\label{energy}
\end{eqnarray}
where the sound velocities $c_\pm$, are defined by \cite{Petrov2}
\begin{eqnarray} 
2c^2_\pm = \sum_{\alpha=\uparrow,\downarrow} g_{\alpha\alpha}n_{\alpha}
\pm \big(\Delta_{gn}^2
+ 4g_{\uparrow\downarrow}^2n_\uparrow n_\downarrow\big)^{1/2}.
\label{sound}
\end{eqnarray}
with $\Delta_{gn}=g_{\uparrow\uparrow}n_{\uparrow} 
- g_{\downarrow\downarrow}n_{\downarrow}$.
In Eq. (\ref{sound}) the sign $+(-)$ corresponds to the two fluids 
moving in phase (out of phase).
The branch $c_-$ could be negative for $g_{11}g_{22}<g_{12}$ indicating 
that the state is unstable and the modes grow exponentially with time
\cite{unstable-modes}.
From Eq. (\ref{energy}), we get the respective $D$-dimensional
chemical potential as
\begin{eqnarray} 
\mu_D (n_{\uparrow},n_{\downarrow}) &=& 
\sum_{\alpha,\sigma=\uparrow,\downarrow}{g_{\alpha\sigma}n_{\alpha}}
+ \frac{\gamma'_D}{L^D}
\big[(g_{\uparrow\uparrow} + g_{\downarrow\downarrow})(c_+ ^D + c_ -^D)
\nonumber \\
&+& (\Delta + \Delta')(c_+ ^D - c_ -^D)\big]
\end{eqnarray}
where $\gamma'_3=5\gamma_3/4$, $\gamma'_1=3\gamma_1/4$,
$ \Delta \equiv [g_{\uparrow\uparrow}
(g_{\uparrow\uparrow}n_\uparrow - g_{\downarrow\downarrow}n_\downarrow)
+ 2g_{\uparrow\downarrow}^2 n_\downarrow]/ \delta $, 
$ \Delta' \equiv [-g_{\downarrow\downarrow}
(g_{\uparrow\uparrow}n_\uparrow - g_{\downarrow\downarrow}n_\downarrow)
+ 2g_{\uparrow\downarrow}^2 n_\uparrow]/ \delta $,
and $\delta \equiv [(g_{\uparrow\uparrow} 
n_\uparrow - g_{\downarrow\downarrow}n_\downarrow)^2
+ 4g_{\uparrow\downarrow}^2 n_\uparrow n_\downarrow]^{1/2} $.
In the case $n_\uparrow\equiv n$, 
$n_\downarrow= 0$, $g_{\uparrow\uparrow}\equiv g$, and 
$g_{\downarrow\downarrow}=g_{\uparrow\downarrow}=0$,
our results are tantamount to those given by the single-component Bose
gases in one dimension \cite{LHY-1D}, and the results in three dimensions
\cite{LHY,Regularization}. 

Recently it has been showed that at zero temperature the quantum 
fluctuations can stabilize the Bose-Bose mixtures and for a finite 
particle number the system gets into a droplet phase 
\cite{Petrov1,Petrov2}.
In order to study not only droplets but also quasi-1D BECs and 3D BECs 
beyond mean-field approximation we consider the case of equal 
intra-species coupling constants 
$g_{\uparrow\uparrow}=g_{\downarrow\downarrow}\equiv g$.
We also consider a symmetric configuration with 
$n_{\uparrow}=n_{\downarrow}\equiv n/2$, where the two internal states are
equally populated. 
In the following analysis we discuss the regime of repulsive intra- and
attractive inter-species interactions. We introduce the relation between 
coupling constants as $\delta g \equiv g - |g_{12}| > 0$ 
with $\delta g \ll g$.
This parameter makes use of the fact that experimental setups of mixtures
of hyperfine states of bosonic alkali atoms are close to the separation 
instability. 
The zero-temperature number density of the droplets $n_D$ can be derived
from the thermodynamic relation $n(\mu)=\partial P/\partial \mu$.
By using the pressures given by Eqs. (\ref{Pressure0}) and
(\ref{Pressure1}), we have
\begin{eqnarray} 
n_D(\mu)= \frac{2\mu}{\delta g} 
-\frac{2\gamma'_D}{L^D} 
\Big[1+\Big(\frac{2g}{\delta g}\Big)^{D/2+1}\Big]\mu^{D/2}.
\end{eqnarray}
In order to study the density profile of a finite number of particles in
the droplet state, we describe the mixture within an effective low-energy 
theory. We use the local density approximation, which leads us to obtain
a zero-temperature $D$-dimensional NLSE of the form
\begin{eqnarray} 
i\hbar \partial_t \psi
&=& -\frac{\hbar^2}{2m}  \partial_D^2 \psi 
+ \frac{1}{2}\delta g |\psi|^2 \psi 
+ 2\frac{\gamma'_D}{L^D}g_D|\psi|^D\psi,
\label{NLSE}
\end{eqnarray}
where $g_D \equiv g^{D/2+1}[1+(\delta g/2g)^{D/2+1}]$, the respective 
$D$-dimensional operator $\partial_D^2$, and the conditions of 
normalization $\int_{-\infty}^{+\infty}d^3 r|\psi(r,t)|^2 = N$, and
$\int_{-\infty}^{+\infty}dx|\psi(x,t)|^2 = N$.
In the particular case $D=1$, the above quadratic-cubic NLSE can support
spatially localized solutions in an explicit analytical form. 
These solutions preserve their shape during the propagation as
spatially self-trapped solitons.
Due to the solitonic nature of these solutions, we name these self-bound
solitonic droplets. Then we sought solutions of the form 
$\psi(x,t)=\phi(x,\mu)\exp(-i\mu t/\hbar)$, with $\phi(x,\mu)$ a real 
function and conditions $\psi = 0$ and 
$d\psi/dx = 0$ as $|x|\rightarrow 0$ \cite{solitons}. 
Provided that $\mu<0$, we found that $\psi(x,t)$ may give rise to 
self-bound brightlike solitonic droplets,
\begin{eqnarray} 
\psi(x,t)= \frac{A\sqrt{n_0}\exp{(i|\mu|t/\hbar})}
{1+\sqrt{1- 3A/4}\cosh{Bx}},
\label{soliton}
\end{eqnarray}
where $A=3|\mu|/2|\mu_0|$, and $B=\sqrt{2m|\mu|}/\hbar$.
The equilibrium density 
$n_0 = mg_1^3/(\pi \hbar \delta g)^2$ is defined by the condition
$\partial\mu/\partial n =0$ and 
$\mu_0 = \mu(n_0)=- mg_1^3/(2\pi^2 \hbar^2 \delta g)$.
From Eq. (\ref{soliton}), brightlike solitonic droplets are present if
$|\mu|< 8|\mu_0|/9$.
So this kind of droplet exists in a finite band $8\mu_0/9 < \mu < 0$, 
rather than in the entire semi-infinite band $\mu<0$. 
When $(\delta g/2g)^{3/2} \rightarrow 0$ in $g_D$, then 
$g_1 \approx g^{3/2}$ and 
the droplet profile given by Eq. (\ref{soliton}) has the same structure to
that previously obtained in Ref. \cite{Petrov2}, where the symmetric case
$n_\uparrow=n_\downarrow\equiv n$ was considered. 
When one of the nonlinear terms in the quadratic-cubic NLSE vanishes, the
solution describes solitons associated with the power-law nonlinearity. 
For example, when $\delta g = 0$ the droplet solution reduces to the
brightlike soliton of the form 
$\psi(x,t)\sim\sech^2(\sqrt{m|\mu|/2\hbar^2}x) \exp{(i|\mu|t/\hbar})$.

Now, we study the NLSE for a very weak attractive value of $\delta g$, 
i.e., $\delta g \lesssim 0$ with $|g_{12}|\gtrsim g$ to ensure the 
condition $|\delta g| \ll g$, such that $g_1=g^{3/2}$.
In that case, we have an instability in the fact that $E_-(k)$ becomes
complex for small momenta $k\sim \sqrt{m|\delta g|n}$.
However, in the most contributing region to the LHY term, i.e., for
$k\gg \sqrt{m|\delta g|n}$ \cite{Petrov1}, both modes $E_\pm (k)$ are not 
affected by the sign of small values of $\delta g$.
So provided that $\mu<0$, the quadratic-cubic NLSE also admits a different
kind of self-trapped darklike solitons of the form
\begin{eqnarray} 
\psi(x,t)= \frac{C\exp{(i|\mu|t/\hbar})} {1 - \sqrt{1+ D} \cosh{Bx}},
\label{soliton2}
\end{eqnarray}
where $C=3|\mu|/\sqrt{2|\delta g|}|\mu'|$, $D=9|\mu|/8|\mu'|$,
and $|\mu'|= mg^3/(2\pi^2 \hbar^2 |\delta g|)$.
Starting from this solution an algebraic soliton $\psi_{\mathrm{alg}}(x)$
can be obtained in the limit $|\mu|\rightarrow0$.
The amplitude of such kind of solitons decays as a power law of the form 
$\phi(x)\sim|x|^{-2}$ for $|x|\rightarrow \infty$ 
\cite{solitons,solitons2}. So, the soliton (\ref{soliton2}) reduces to 
\begin{eqnarray} 
\psi_\mathrm{alg}(x)= - \frac{3\hbar^2}{\sqrt{2|\delta g|}m|\mu'|}
\frac{1}{x^2 + 9\hbar^2/16m|\mu'|}.
\label{alg}
\end{eqnarray}
Numerically, these algebraic solitons belong to a branch of the unstable 
sechlike solitons \cite{solitons2}. 
As a result of this instability, an initially perturbed algebraic soliton
undergoes a switching to a stable branch of the sech-type solitons and it
is accompanied by the large-amplitude oscillations.

Regarding the applicability of the 1D NLSE, note that, in order to justify
the inclusion of the LHY term as a local contribution, the distances at 
which the density changes should be large compared to the healing length 
\cite{1D-droplet}.
The density changes at distances $\xi_-= \hbar/(\sqrt{2}mc_-)$, where
$c_-=n\delta g/2$, while the LHY term is considered at distances 
comparable to $\xi_+= \hbar/(\sqrt{2}mc_+)$, with $c_+=gn$.
As $g\gg\delta g$, the large separation of scales, $\xi_+ \ll\xi_-$, 
justifies the use of the LHY term as a local contribution in the 1D NLSE.

\section{Finite-temperature results}
In the continuum limit, the one-loop contribution to the grand potential
at finite temperature is obtained from the second term of 
Eq. (\ref{Grand2}), and it is written as
\begin{eqnarray} 
L^{-D}\Omega_{g}^{(T)}(\mu_{\uparrow},\mu_{\downarrow})
= A_D \sum_{\pm}\int_0^{+\infty} dE_{\pm}(k)
\frac{k^{D}_{\pm}}{e^{\beta E_{\pm}(k)}-1},
\end{eqnarray}
where $A_3=-(6\pi^2)^{-1}$ and $A_1=-\pi^{-1}$.
Introducing the variable $x_{\pm}=\beta E_{\pm}$, we get
\begin{eqnarray} 
L^{-D}\Omega_{g}^{(T)}(\mu_{\uparrow},\mu_{\downarrow})
= \frac{A_D}{\beta} 
\sum_{\pm}\int_0^{+\infty} dx_{\pm}
\frac{k^{D}_{\pm} (x_{\pm},f_{\pm},T)}{e^{x_{\pm}}-1},
\label{grandT}
\end{eqnarray}
where $k_{\pm}^2 (x_{\pm},f_{\pm},T)$ is given by
\begin{eqnarray} 
k^2_{\pm} (x_{\pm},f_{\pm},T) = \frac{2mf^2_{\pm}}{\hbar^2}
\Bigg(\sqrt{1+\frac{x_{\pm}^2}{\beta^2 f_{\pm}^4}} -1 \Bigg).
\end{eqnarray}
The thermal contribution to $\Omega^{(T)}_{g}$ is affected by two-body 
interactions through the dependence of the Bogoliubov modes $E_{\pm}(k)$ 
on the intra- and inter-species interaction coupling
constants $g_{\alpha\alpha}$ and $g_{\alpha\sigma}$ \cite{Pit-Strin}.
The one-loop contribution to the equation of state of $D$-dimensional Bose-Bose
gases at low temperature can be calculated by means of the expansion of the
Eq. (\ref{grandT}).
This integral yields the result
\begin{eqnarray} 
P^{(T)}_{D}&=& -A_D(k_B T)^{D+1} \sum_{\pm}
\Big(\frac{m}{\hbar^2f^2_{\pm}}\Big)^{D/2}
\Big[\Gamma(D+1)\zeta(D+1) 
\nonumber \\
&-& \frac{D}{8}\Big(\frac{k_BT}{f^2_{\pm}}\Big)^2 
\Gamma(D+3)\zeta(D+3)\Big]
\end{eqnarray}
holding for $k_B T\ll mc^2_{\pm}$ \cite{Pit-Strin}.
In a very weak-interacting Bose-Bose mixture at finite temperature, the 
contribution of $f_\pm$ increases the role of the high corrections to the 
equation of state. This is achieved providing that both the intra- and the
inter-species coupling constants tend to zero. 
Now, we derive the thermal equation of state for symmetric droplets. We 
assume $g_{\uparrow\uparrow}=g_{\downarrow\downarrow}\equiv g$,
$n_{\uparrow}=n_{\downarrow}\equiv n/2$ with 
$\delta g \equiv g - |g_{12}| > 0$ and $\delta g \ll g$. Thus
\begin{eqnarray} 
P^{(T)}_{D}&=& -A_D(k_B T)^{D+1} \Big(\frac{m}{\hbar^2\mu}\Big)^{D/2}
\Big[\epsilon_1 \Gamma(D+1)\zeta(D+1) 
\nonumber \\
&-& \epsilon_2 \frac{D}{8} \Big(\frac{k_BT}{\mu}\Big)^2 
\Gamma(D+3)\zeta(D+3)\Big],
\label{P_drop}
\end{eqnarray}
where $\epsilon_1=[1+(\delta g/2g)^{D/2}]$, and
$\epsilon_2=[1+(\delta g/2g)^{D/2 +2}]$. 
Equation (\ref{P_drop}) generalizes some zero-temperature 
results achieved in Refs. \cite{Petrov1,Petrov2}. We also obtain the
number density of the quantum droplets at finite temperature,
\begin{eqnarray} 
n_{D}^{(T)}
&=& A_D(k_B T)^{D+1} \frac{D}{2\mu} \Big(\frac{m}{\hbar^2\mu}\Big)^{D/2}
\Big[\epsilon_1\Gamma(D+1)\zeta(D+1) 
\nonumber \\
&-&
\frac{\epsilon_2}{4}\Big(\frac{D}{2}+2\Big)\Big(\frac{k_B T}{\mu}\Big)^2
\Gamma(D+3)\zeta(D+3)\Big].
\end{eqnarray}

\section{Conclusion}
We derive the equation of state at zero and finite temperature for one- and
three-dimensional dilute and ultracold
Bose-Bose mixtures of alkali atoms. We perform one-loop functional integration
and the ultraviolet divergent zero-point energy is removed through
dimensional regularization.
Our findings also show that the 1D NLSE to describe Bose-Bose gases
supports analytical self-trapped brightlike solitonic droplets and 
self-trapped darklike solitons.
Our analytical results at finite temperature are a nontrivial 
generalization of those obtained by using the standard Bogoliubov theory
at zero temperature \cite{Petrov1,Petrov2}.
We believe that our theoretical predictions could stimulate interesting 
experimental work in the study of beyond-mean-field effects in Bose-Bose
mixtures.
As an extension of the present work, it may be interesting to verify the 
validity of our results by means of the quantum Monte Carlo technique.
Other important questions, such as those that consider the finite-range
effects of the interatomic potential, remain to be investigated in the 
future.

\end{document}